\documentclass[journal]{IEEEtran}
\usepackage{upgreek}
\usepackage{graphicx}
\usepackage{subfigure}
\usepackage{amsmath,bm}
\usepackage{enumerate}

\usepackage[square, comma, sort&compress, numbers]{natbib}

\usepackage{color}

\ifCLASSINFOpdf
\else
\fi

\begin{document}

\title{\huge Mobile Optical Communications \\ Using Second Harmonic of Intra-Cavity Laser}
\author{
	Mingliang Xiong, Qingwen Liu*,~\IEEEmembership{Senior Member,~IEEE}, Xin Wang,~\IEEEmembership{Senior Member,~IEEE},\\ Shengli Zhou,~\IEEEmembership{Fellow,~IEEE}, Bin Zhou, and Zhiyong Bu

\thanks{
	* The corresponding author: Qingwen Liu.
}
\thanks{
	M. Xiong and Q. Liu
	are with the College of Electronics and Information Engineering, Tongji University, Shanghai 201804, China
	(e-mail: xiongml@tongji.edu.cn;    qliu@tongji.edu.cn).
	X. Wang is with the Key Laboratory for Information Science of Electromagnetic Waves (MoE), Department of Communication Science and Engineering, Fudan University, Shanghai 200433, China (e-mail: xwang11@fudan.edu.cn). S. Zhou is with Department of Electrical and Computer Engineering, University of Connecticut, Storrs, CT 06250, USA (e-mail:
	shengli.zhou@uconn.edu). B. Zhou and Z. Bu are with the Key Laboratory of Wireless Sensor Network and Communications, Shanghai Institute of Microsystem and Information Technology, Chinese Academy of Sciences, 865 Changning Road, 	Shanghai 200050, China (email: bin.zhou@mail.sim.ac.cn, zhiyong.bu@mail.sim.ac.cn).
	}

\thanks{This work was supported by Shanghai Municipal Science and Technology Major Project (2021SHZDZX0100) and Shanghai Municipal Commission of Science and Technology Project (19511132101). It was also supported by the National Natural Science Foundation of China under Grant 62071334 and Grant 61771344, and the National Key Research and Development	Project under Grant 2020YFB2103902.}
}

\maketitle

\begin{abstract}
Optical wireless communication~(OWC) meets the demands of the future six-generation mobile network (6G) as it operates at several hundreds of Terahertz and has the potential to enable data rate in the order of Tbps. However, most beam-steering OWC technologies require high-accuracy positioning and high-speed control. Resonant beam communication~(RBCom), as one kind of non-positioning OWC technologies, has been proposed for high-rate mobile communications. The mobility of RBCom relies on its self-alignment characteristic where no positioning is required. In  a previous study, an external-cavity second-harmonic-generation (SHG) RBCom system has been proposed for eliminating the echo interference inside the resonator. However, its energy conversion efficiency and complexity are of concern. In this paper, we propose an intra-cavity SHG RBCom system to simplify the system design and improve the energy conversion efficiency. We elaborate the system structure and establish an analytical model. Numerical results show that the energy consumption of the proposed intra-cavity design is reduced to reach the same level of channel capacity
at the receiver compared with the external-cavity one.
\end{abstract}


\section{Introduction}\label{sec:intro}

\begin{figure}[t]
	\centering
	\includegraphics[width=3.2in]{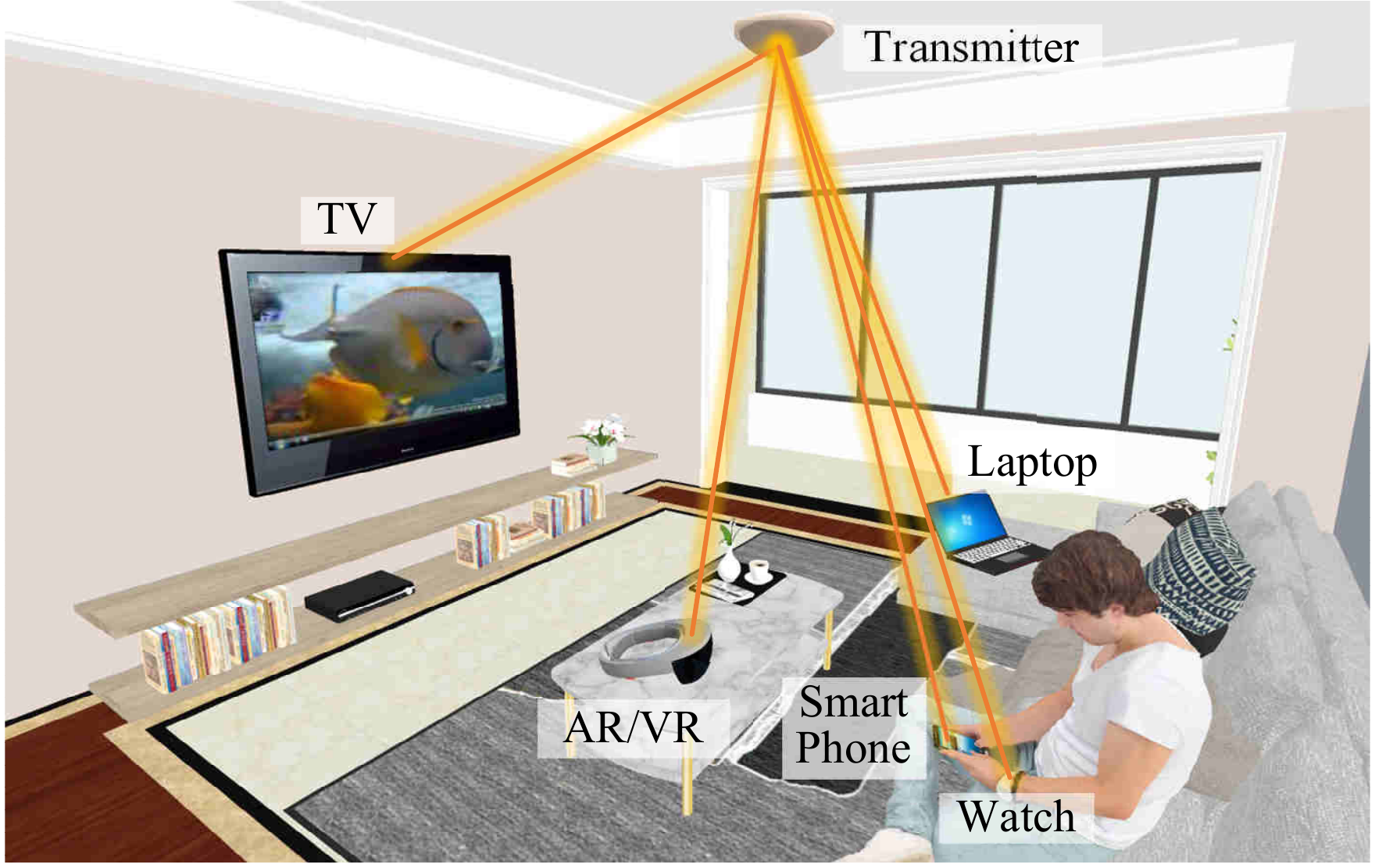}
	\caption{Application scenario of mobile optical communications}
	\label{fig:appli}
\end{figure}

\IEEEPARstart{T}{o} meet the requirements of  future mobile Internet applications, such as holographic and high-precision light field transmission, full sensory (vision, hearing, touch, smell, and taste) perception, etc., the data rate is expected to be over $1$~Tbit/s~\cite{a210529.02, a180820.09}. To achieve this goal, many researchers start to focus on infrared or visible light spectrum, as it can provide very large bandwidth~\cite{a210529.03}. Therefore, optical wireless communication~(OWC) has been widely studied. For example, visible light communication~(VLC), free-space optical communication~(FSO), light fidelity~(Li-Fi), and optical camera communication~(OCC) are well-known OWC technologies~\cite{HLi2020, RLi2018, Krishnan2018,Nguyen2020}. The proposed mobile optical communication (MOC) in this paper belongs to OWC, which adopts a spatially separated laser resonator (SSLR) structure 
to enable mobility and high-rate communication, as shown in  Fig.~\ref{fig:appli}.

\begin{figure}[t]
	\centering
	\includegraphics[width=3.0in]{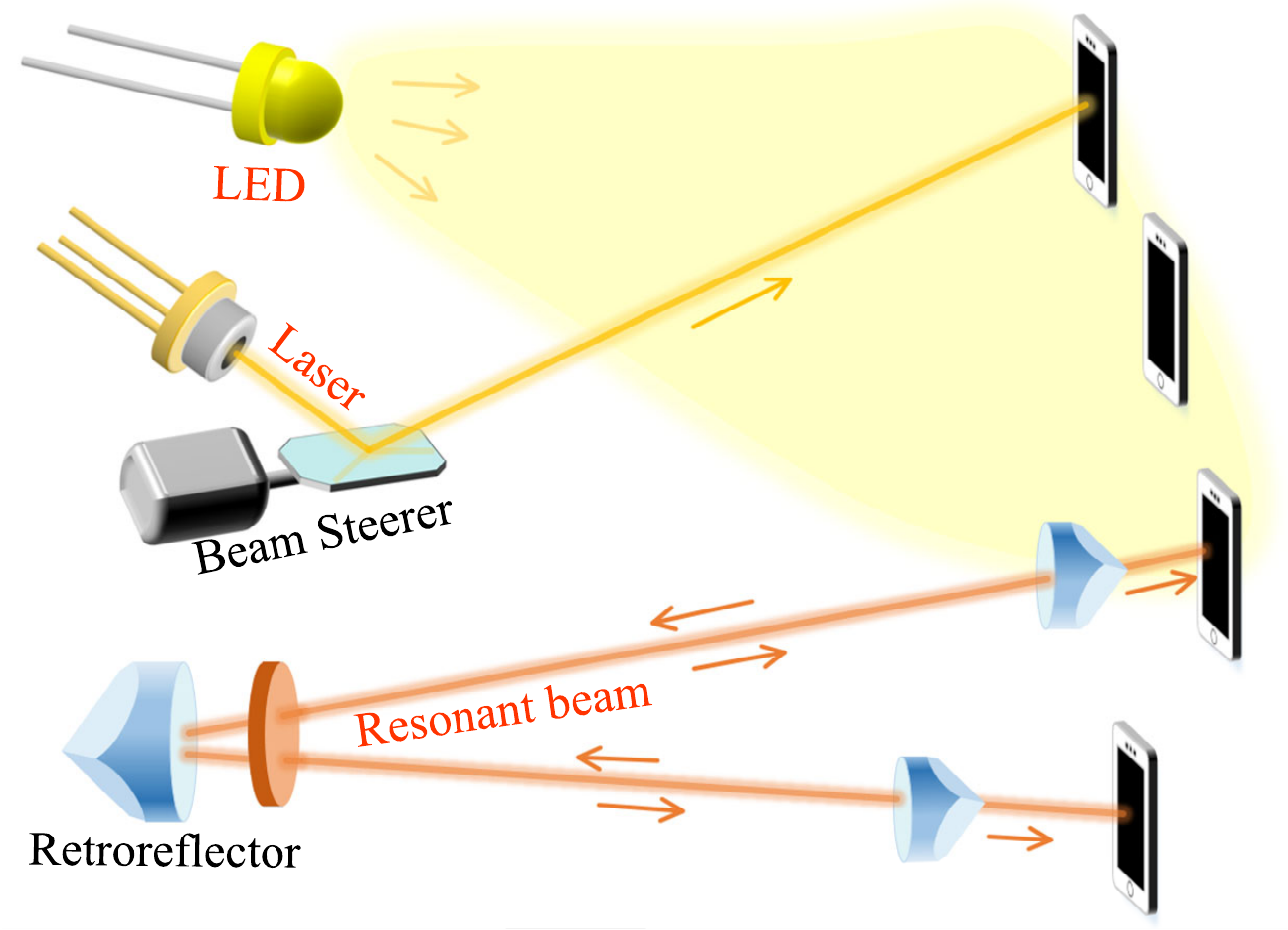}
	\caption{Technologies comparison on optical wireless communications}
	\label{fig:scenario}
\end{figure}

In typical OWC systems, light-emitting diodes~(LEDs) or laser diodes~(LDs) are adopted. As demonstrated in Fig.~\ref{fig:scenario}, 
LED as the transmitter emits light to a large area and the receivers can receive the signal in its coverage. Multiple LEDs mounted on the ceiling can cover a large receiving space~\cite{a190909.52}. However, the signal attenuation of LED radiation is severe due to the path loss over the large divergence angle. To enhance the received power, aspheric condenser lenses were proposed to focus the LED radiation on the receiver, however this scheme is only suitable for point-to-point (P2P) transmission~\cite{a210529.04}. 

To improve the purity of the carrier frequency, LDs become popular in OWC research. Laser generated by stimulated emission has good directivity and coherence for high received power and channel capacity. 
There are many methods to enable mobility in laser OWC. For example, Chun \textit{et. al.} used micro-electromechanical-system (MEMS) mirrors to steer the laser beam to the receiver and achieved $35$~Gbit/s in a wide-area coverage~\cite{a201201.03}. To speed up the response of the beam steerer, many non-mechanical schemes emerge. For instance, Wang \textit{et. al.} exploited in-fiber diffraction grating as the beam steerer and demonstrated $12$-Gbit/s-per-beam transmission~\cite{a201130.03}. To realize two-dimensional~(2D) beam steering, a basic idea is to use an assembled 2D fiber array and a lens. Each fiber is in response to a specified small area in the receiving plane. The lens casts the image of the array plane to the receiving plane. This work was well demonstrated in~\cite{a191111.01}. Silicon optical phased arrays (OPA) comprised of electro-optic phase shifters and thermo-optic radiators are another kind of devices for beam steering. Rhee \textit{et. al.} employed the OPA as the beam steerer and achieved $32$-Gbit/s data transmission with the coverage of $46.0^\circ \times 10.2^\circ$ in two orthogonal directions~\cite{a201201.02}. Gratings can also be assembled for 2D beam steering. Koonen \textit{et. al.} achieved transmitting $32$~Gbit/s via an infrared beam over a $6^\circ\times12^\circ$ area by a pair of crossed gratings~\cite{a190611.03}. Besides P2P transmission, point-to-multipoint beam steering also received great attention. For example, Gomez \textit{et. al.} used spatial light modulator~(SLM) as the beam steerer and demonstrated a $50$-Gbit/s optical link with $\pm30^\circ$ field-of-view~\cite{a190514.01}. Zhang \textit{et. al.} verified the mobile communication performance of the system using a SLM for multiple-beam steering. Despite these advantages, OWC technologies still have the challenges in receiver positioning and tracking. For a narrow beam, the positioning accuracy and control latency are the dominant limitations for tracking a mobile receiver, especially over a long distance.

As shown in Fig.~\ref{fig:scenario}, we propose a resonant beam communication~(RBCom) system based on SSLR and second harmonic generation~(SHG) to provide a non-positioning and high-capacity mobile optical link. The primary idea of SSLR was proposed by Linford \textit{et. al.} in 1973, as a very long laser resonator, for atmospheric pollution detection~\cite{a190318.02}. Basically, the SSLR comprised of two retroreflectors, such as corner cubes or cat's eyes. Photons oscillate in the resonator and are amplified by the gain medium to form an intra-cavity resonant beam. Therefore, it is possible to transmit information by modulating signal on the resonant beam. Liu \textit{et. al.} presented the application of charging mobile electronics using SSLR, which is also known as distributed laser charging~(DLC)~\cite{a180727.01}.  DLC with spatial wavelength division was well demonstrated in~\cite{a190926.02}. The application in simultaneous wireless information and power transfer~(SWIPT) was presented in  \cite{a210601.01}. The self-aligned mobility and the safety of the SSLR for high-power transmission were verified in~\cite{MLiu2021} and~\cite{WFang2021}, respectively. The application of communications using SSLR was illustrated in~\cite{a191111.06}. Since the resonant beam oscillates in the cavity, the modulated signal will also be reflected by the retroreflectors, which may result in echo interference. In order to eliminate or avoid the echo interference, several methods were proposed. In \cite{MXiong2021}, an external-cavity SHG crystal was employed for changing the carrier frequency to its doubled value and thus the echo interference can be avoided. The frequency-doubled carrier is filtered at the receiver out of the resonator. Due to the fact that the beam put into the SHG crystal is extracted by a splitter from the resonant beam, energy conversion is not efficient. To address this issue, we propose an intra-cavity SHG design in this paper. 

The contributions of our work are as follows.
\begin{enumerate}
	\item[\bf 1)] We propose an intra-cavity SHG scheme for the SSLR-based communication system to avoid the echo interference, which enables simplified system structure and higher energy conversion efficiency compared with the previous external-cavity SHG scheme.
	
	\item[\bf 2)] We design the system structure which provides a shared optical path for both the fundamental resonance and the information carrier. Therefore, the information carrier can be aligned with the receiver automatically without any beam steering devices, which enables the  mobility. This structure also offers a fixed small modulation spot for the carrier beam so that the energy efficiency and modulation speed can be improved.
	
	\item[\bf 3)] We develop an analytical model and evaluate the system performance including the beam  radius, the received power, and the channel capacity.
	
\end{enumerate}

The remainder of this paper is organized as follows.
In Section~\ref{sec:design}, we propose a new system design. In Section~\ref{sec:model}, we present the analytical model. In Section~\ref{sec:result}, we demonstrate the system performance in terms of the beam radius, the received power, and the channel capacity. Finally, we make conclusions in Section~\ref{sec:con}.

\section{The Proposed System}
\label{sec:design}

	SHG crystals are devices to perform frequency doubling.
	Here, we term the input to the SHG crystal as the fundamental beam, and the output as the frequency-doubled beam. In~\cite{MXiong2021}, 
	we proposed a system, where part of the resonant beam in the cavity is extracted and used as the fundamental beam to the SHG crystal. The SHG crystal is placed outside the laser resonator. In this paper, we place the SHG crystal inside the resonator and hence the resonant beam in the cavity itself is used as the fundamental beam. Following this change, we modify the structure.
	
	\begin{figure*}[t]
		\centering
		\includegraphics[width=5.4in]{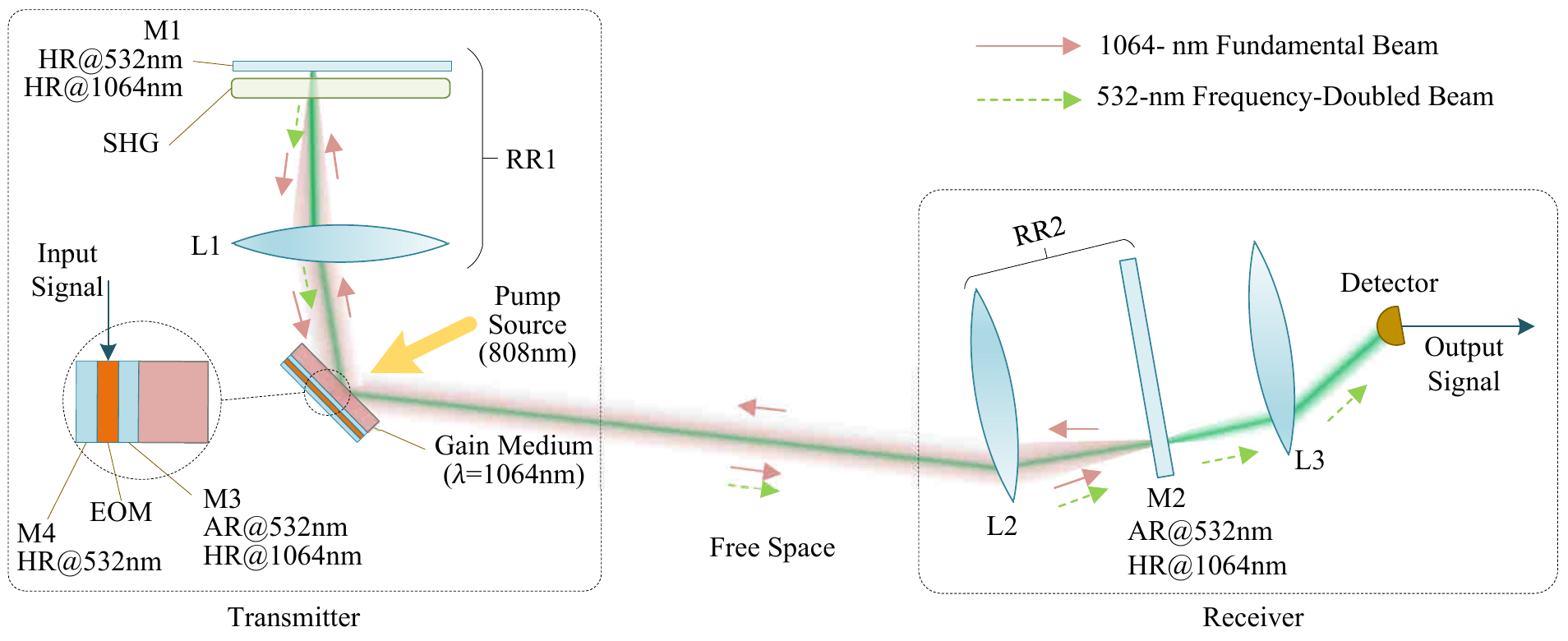}
		\caption{Design of the intra-cavity second harmonic generation-based resonant beam communication system (M1, M2 are mirrors; M3 and M4 are coatings; L1 to L3 are lenses; RR1 is the retroreflector at the transmitter, consisting of M1 and L1; RR2 is the retroreflector at the receiver, consisting of M2 and L2; EOM is the electro-optical modulator; AR: anti-reflective coating, i.e., transmissivity $\approx 100$\%; HR: high-reflective coating, i.e., reflectivity $\approx 100$\%) }
		\label{fig:design}
	\end{figure*}

	The system structure includes an SSLR, an SHG crystal, and some modulating and detecting devices. Different from conventional laser resonant cavities  comprised of two mirrors, the SSLR consists of two telecentric cat's eye retroreflectors RR1 and RR2 which integrate a lens and a rear mirror placed behind the lens with a separation. One of the retroreflector is located at the transmitter, the other one is located at the receiver, as shown in Fig.~\ref{fig:design}. At the transmitter, a laser gain medium is placed at the pupil of the retroreflector RR1. The resonance initiates in the SSLR when certain conditions are satisfied; and then, the resonant beam is generated inside the cavity. The SHG crystal is placed inside RR1 to generate frequency-doubled beam from the resonant beam (which is called the fundamental beam when referring to SHG). The frequency-doubled beam is transmitted via the optical path of the fundamental beam, i.e., it passes the pupil of RR1 and then is received by RR2.
		
\begin{figure}[t]
	\centering
	\includegraphics[width=3.2in]{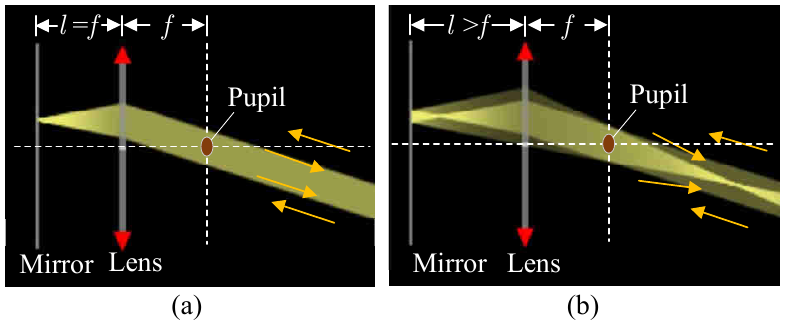}
	\caption{Telecentric cat's eye retroreflector structure in two cases: (a) The interval between the lens and the mirror $l$ $=$ the focal length of the lens $f$; and (b) $l>f$}
	\label{fig:retro}
\end{figure}

		\begin{table} [t]
			\caption{Transfer Matrices of Elementary Optical Structures}
			\renewcommand{\arraystretch}{1.2}
			\centering
			\begin{tabular}{|c|c |c|}
				\hline
				\textbf{Optical Element}&\textbf{Diagram} 
				& \textbf{ Transfer Matrix}  \\
				
				\hline
				Free Space&
				\begin{minipage}[h]{0.3\columnwidth}
					\centering
					\includegraphics[width=\linewidth]{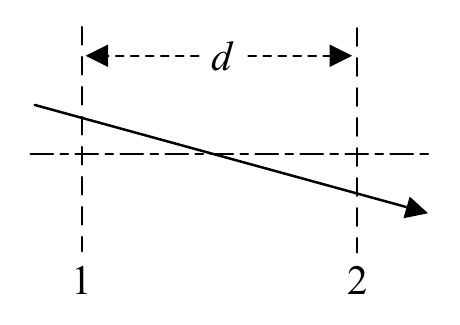}
				\end{minipage}
				& 
				\begin{minipage}[h]{0.2\columnwidth}
					\centering
					$
					\begin{bmatrix}  ~1 & d~ \\ ~0 & 1~ \end{bmatrix}
					$
				\end{minipage}
				\\
				\hline
				Lens&
				\begin{minipage}[h]{0.3\columnwidth}
					\centering
					\includegraphics[width=\linewidth]{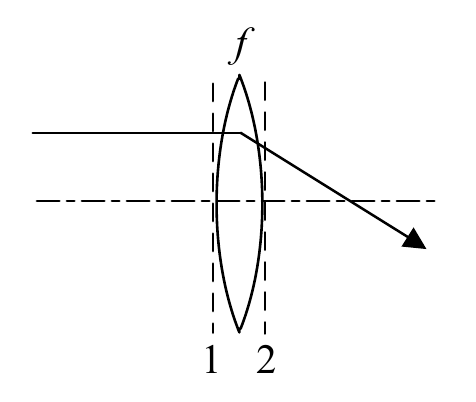}
				\end{minipage}
				& 
				\begin{minipage}[h]{0.2\columnwidth}
					\centering
					$
					\begin{bmatrix}  ~1 & 0~ \\ ~-1/f & 1~ \end{bmatrix}
					$
				\end{minipage}
				\\
				\hline
				Mirror&
				\begin{minipage}[h]{0.3\columnwidth}
					\centering
					\includegraphics[width=\linewidth]{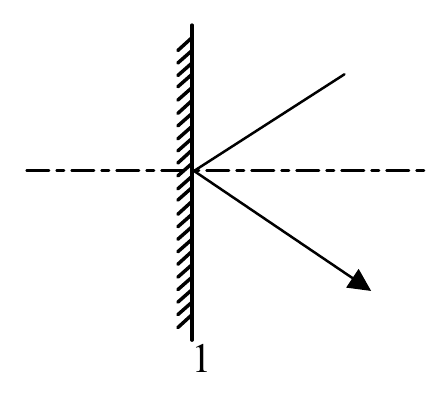}
				\end{minipage}
				& 
				\begin{minipage}[h]{0.2\columnwidth}
					\centering
					$
					\begin{bmatrix}  ~1 & 0~ \\ ~0 & 1~ \end{bmatrix}
					$
				\end{minipage}
				\\
				\hline
			\end{tabular}
			\label{tab:optmatrix}
		\end{table}

		The pupil of a telecentric cat's eye retroreflector is a special spot. Beams from any directions that pass this spot can be reflected to their incoming directions by the retroreflector, as depicted in Fig.~\ref{fig:retro}. This function supports the mobility of the SSLR, as photons can be captured to oscillate between two arbitrarily located retroreflectors. The telecentric cat's eye retroreflector is comprised of a mirror and a lens. The interval, $l$, between the mirror and the lens decides how the reflected beam is focused. In geometrical optics, an optical system can be expressed by a ray transfer matrix. The matrices of elementary optical devices are listed in Table~\ref{tab:optmatrix}. The matrix of other complex system that consists of these elements can be expressed by the product of the matrices of the elements in this system in opposite order. Namely, the matrix of the device that is crossed by the ray first is placed to the right in the formula. For a telecentric cat's eye retroreflector whose  internal lens has focal length of $f$, the transfer matrix can be obtained as~\cite{MXiong2021}
		\begin{align}
			\mathbf{M}_{\rm RR}&=			\begin{bmatrix}
			1&f\\0& 1
			\end{bmatrix}
			\begin{bmatrix}
			1&0\\-1/f& 1
			\end{bmatrix}
			\begin{bmatrix}
			1&l\\0& 1
			\end{bmatrix}
			\begin{bmatrix}
			1&0\\0& 1
			\end{bmatrix}
			\begin{bmatrix}
			1&l\\0& 1
			\end{bmatrix}
			\nonumber
			\\
			&~~~~
			\begin{bmatrix}
			1&0\\-1/f& 1
			\end{bmatrix}
			\begin{bmatrix}
			1&f\\0& 1
			\end{bmatrix}
			\nonumber
			\\
			&=
			\begin{bmatrix}
			1&0\\-1/f_{\rm RR}& 1
			\end{bmatrix}
			\begin{bmatrix}
			-1&0\\0 &-1
			\end{bmatrix},
		\end{align}
		where
		\begin{equation}
		f_{\rm RR}=\dfrac{f^2}{2(l-f)}.
		\label{equ:fRR}
		\end{equation}
	It can be found that $\mathbf{M}_{\rm RR}=-\mathbf{I}$, where $\mathbf{I}$ is $2\times 2$ unit matrix, if $l=f$. In this case, the matrix represents an ideal telecentric cat's eye which can reflect the ray back along the parallel path of the incoming path, as depicted in Fig.~\ref{fig:retro}(a). But, we 
	choose $l>f$ when designing an SSLR, as the retroreflector with this configuration also provides a focusing ability in favor of creating a stable resonator, as shown in Fig.~\ref{fig:retro}(b).

	There are many kinds of laser gain mediums such as crystals, glasses, semiconductors, and gases. They absorb pump light at a frequency and then amplify light at another lower frequency. Different materials determine which frequency to be absorbed and amplified. For our system, the gain medium to be used should be transparent at the second-harmonic frequency. In this paper, we employ neodymium-doped yttrium orthovanadate (Nd:YVO$_4$) which absorb $808$-nm light and amplify $1064$-nm light. Hence, the frequency-doubled beam generated by the SHG crystal is at $532$~nm. The selected frequencies of the fundamental beam and the frequency-doubled beam lie in the atmospheric window, i.e., $400-1300$~nm, so that they experience less propagation loss. The gain medium is placed at the pupil of RR1 so that the intra-cavity resonant beam always passes through the gain medium. This configuration reduces the area of the gain medium and ensures that the pump light can be focused onto a fixed small spot. A gain medium with a small cross section area exhibits a small threshold power and therefore a high end-to-end efficiency, which increases the pump energy utilization.
	
	Behind the gain medium, there is a coating M3 which is high-reflective (HR, reflectivity $\approx 100\%$ ) at $1064$ nm and anti-reflective (AR, transmissivity $\approx$~100\%) at $532$~nm. Therefore, the $532$-nm beam can pass through the gain medium and then reach the electro-optical modulator~(EOM). After being modulated by the EOM, the $532$-nm beam is reflected by the coating M4. Since the coating and the EOM are very thin, the $532$-nm beam can still transit along the path of the $1064$-nm fundamental beam. Nanoparticle film modulator or pixelated multiple quantum-well~(MQW) modulator can be employed here~\cite{a190909.23,a190926.04}, as they are thin enough. Different from the previous work, here we place the EOM at the pupil of RR1 so that the area of the EOM can be small. This can improve the modulation speed and reduce the energy consumption~\cite{a200512.02}. In this design, the pump beam incidents on the AR coated surface of the gain medium with 0$^\circ$ incidence angle. On the other side, the HR-coated surface of the EOM is attached to a heat sink. Then, we can use circulating water or semiconductor thermoelectric  cooler to take away the heat.
	
	At the receiver, the mirror M2 reflects the fundamental beam to maintain the resonance, and only allows the frequency-doubled beam to pass. In both RR1 and RR2, the beam paths are perpendicular to the rear mirror, which satisfies the incident angle requirement of the SHG crystal, and also ensures that the beam can always be focused onto the photodetector~(PD) by lens L3. With the proposed system structure, the frequency-doubled beam is modulated by the EOM and transits along the path of the fundamental beam pointing to the receiver.
	
\section{System Analysis}	
\label{sec:model}	
\subsection{Fundamental Beam Generation}

\begin{figure}[t]
	\centering
	\includegraphics[width=3.4in]{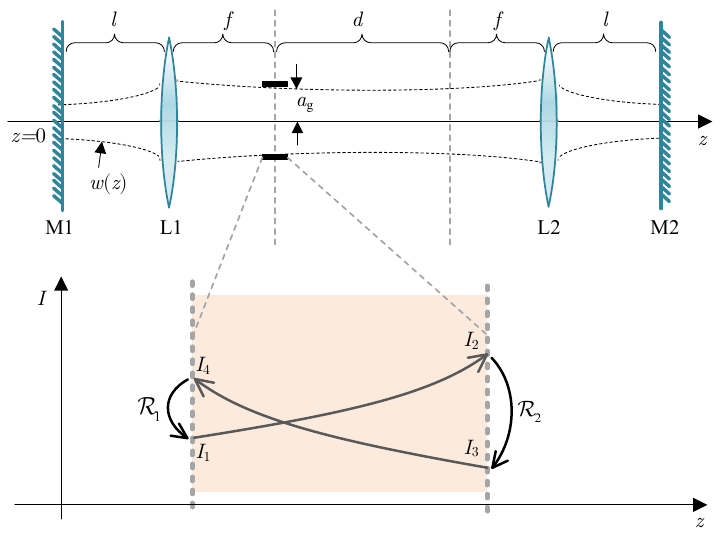}
	\caption{Intra-cavity beam distribution and circular power model}
	\label{fig:model}
\end{figure}

	Now, we illustrate the model of the intra-cavity fundamental beam generated by the SSLR. The analysis includes the stability of a resonator and the mathematic formula for calculating the power of the fundamental beam. 
	
	A stable resonator means that light rays on-axis will remain in it after many bounces. As demonstrated in Fig.~\ref{fig:model}, the SSLR is comprised of mirrors M1 and M2, and lenses L1 and L2. In order to analyze the stability of the SSLR, we write down its signal-pass ray-transfer matrix (also be called as ABCD matrix)~\cite{a200224.01,a190511.01}; that is 
	\begin{align}
	\begin{bmatrix}
	A&B\\C&D
	\end{bmatrix}
	=&	\begin{bmatrix}
	1&0\\0&1
	\end{bmatrix}
		\begin{bmatrix}
	1&l\\0&1
	\end{bmatrix}
		\begin{bmatrix}
	1&0\\ -1/f&1
	\end{bmatrix}
		\begin{bmatrix}
	1&2f+d\\0&1
	\end{bmatrix}	
	\nonumber
	\\
		&\begin{bmatrix}
	1&0\\ -1/f&1
	\end{bmatrix}
		\begin{bmatrix}
	1&l\\0&1
	\end{bmatrix}
		\begin{bmatrix}
	1&0\\0&1
	\end{bmatrix}.
	\end{align}
	We obtain
	\begin{align}
	\left\{
	\begin{array}{lr}
	A=-1-\dfrac{d}{f}+\dfrac{dl}{f^2}\\
	B=2f-2l+d-\dfrac{2dl}{f}+\dfrac{dl^2}{f^2}\\
	C=\dfrac{d}{f^2}\\
	D=-1-\dfrac{d}{f}+\dfrac{dl}{f^2}\\
	\end{array}
	\right.
	\label{equ:ABCD}
	\end{align}
	We set $g_1^*=A$, $g_2^*=D$, and $L^*=B$, and use these parameters to analyze whether the resonator is stable and calculate the beam radius. Referring to \cite{a181221.01}, for a stable resonator, the condition $0<g_1^*g_2^*<1$ must be satisfied. Using $f_{\rm RR}$ defined by~\eqref{equ:fRR} to replace the parameter $l$ and $f$ in~\eqref{equ:ABCD}, we can derive that $0\le d <4f_{\rm RR}$ should be satisfied for a stable SSLR~\cite{MXiong2021}.
	
	The $g*$ parameters can also be used to calculate the beam radius. The intra-cavity fundamental beam consists of multiple transverse modes. Among these modes, the fundamental mode TEM$_{00}$ has the smallest radius. Note that the fundamental mode is different from the aforementioned fundamental beam. We only use the term fundamental beam when  referring to the SHG process. The intensity of the fundamental mode exhibits a Gaussian distribution in transverse plane, and its radius can be computed by
	\begin{equation}
	w_{00}(z)=\sqrt{-\dfrac{\lambda}{\pi\Im\left[1/q(z)\right]}},
	\end{equation}
	where $\Im[\cdot]$ takes the imaginary part of a complex number, and  $\pi$ is the ratio of a circle's circumference to its diameter. The parameter $q(z)$ records all the information of the fundamental mode at location $z$, including the mode's radius and the radius of curvature of the constant-phase surface. Generally, we can use the $g^*$ parameters of the resonator to obtain the $q(z)$ at any location on the $z$-axis of SSLR; the result is from~\cite{MXiong2021}
\begin{equation}
q(z)=
\left\{
\begin{aligned}
&q_0+z, ~~~~~~~~~~~~~~~~~~~~~~~~~~z\in[0,z_{\rm L1}]\\
&\frac{q(z_{\rm L1})}{{-q(z_{\rm L1})}/{f}+1}+(z-z_{\rm L1}), z\in(z_{\rm L1},z_{\rm L2}]\\
&\frac{q(z_{\rm L2})}{{-q(z_{\rm L2})}/{f}+1}+(z-z_{\rm L2}), z\in(z_{\rm L2},z_{\rm M2}]\\
\end{aligned}
\right.
\label{equ:qParam}
\end{equation}
where $z_{\rm L1} = l$, $z_{\rm L2} = l+d+2f $, and $z_{\rm M2} = 2l+d+2f $. From \eqref{equ:qParam}, we can see that all the q-parameters are derived from the q-parameter at $z=0$. At this spacial location, the q-parameter can be computed by~\cite{a181221.01,a200515.04}
\begin{equation}
	q_0=j |L^*|\sqrt{\dfrac{g_2^*}{g_1^*(1-g_1^*g_2^*)}}.
\end{equation}

\begin{figure}[t]
	\centering
	\includegraphics[width=3.5in]{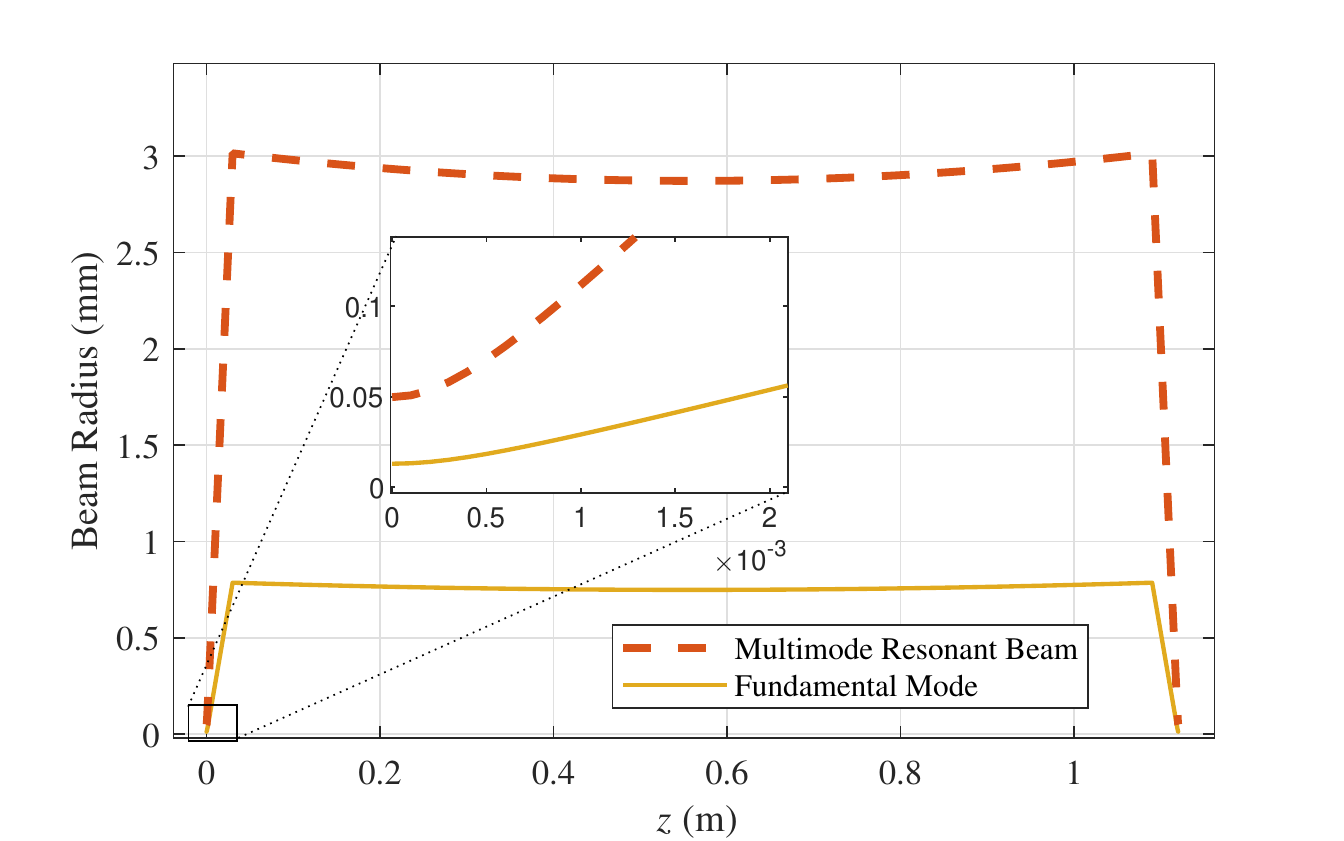}
	\caption{Beam radius distribution along the $z$-axis ($d=1$~m, $f=3$~cm, $l=3.015$~cm)}
	\label{fig:r-z}
\end{figure}

The beam radius of the fundamental beam is approximately proportional to the radius of the TEM$_{00}$ mode. The proportion is also called the beam propagation factor, which is a constant along the $z$-axis. We can use the radii of the fundamental beam and the TEM$_{00}$ mode at the gain medium ($z=l+f$) to compute this beam propagation factor. Since the aperture of the gain medium is the smallest one anomg the devices in the SSLR, the beam raidus at the gain medium can be assumed to be equal to the radius of the gain medium $a_g$. Then, the beam propagation factor can be obtained as $a_{\rm g}/w_{00}(l+f)$. Finally, the radius of the fundamental beam at arbitrary location $z$ can be approximated as~\cite{a181221.01}
\begin{equation}
w(z)=\dfrac{a_{\rm g}}{w_{00}(l+f)}\sqrt{-\dfrac{\lambda}{\pi\Im\left[1/q(z)\right]}}.
\label{equ:radius}
\end{equation}
From \eqref{equ:radius}, we can obtain the beam radius at the gain medium; that is $w(l+f)$. Fig.~\ref{fig:r-z} demonstrates the radii of the multi-mode resonant beam and the fundamental mode distributed along the $z$-axis, where we set $f=3$~cm, $l=3.015$~cm, and $d=1$~m. We can see that the beam radius between L1 and L2 is approximate to $3$~mm; and the beam waist locates at half of the transmission distance. Between the lens and the mirror, the beam is focused by the lens and reaches the minimum radius at the mirror. In addition, we can observe that the difference of the beam radius in the region close to the mirror is small.

To obtain the power of the fundamental beam, we also need to calculate the light intensity of the fundamental beam at the gain medium. As demonstrated in Fig.~\ref{fig:model}, the SSLR is equivalent to a simple resonator which consists of two mirrors and a gain medium placed between them. The reflectivities of the two mirrors  $\mathcal{R}_1$ and $\mathcal{R}_2$ represent the combined losses at the left-side and the right-side of the gain medium, respectively. We use $I_4$ to denote the beam intensity at the point where the beam incidents the mirror $\mathcal{R}_1$. According to the Rigrod analysis, given the source power $P_{\rm in}$, we can obtain~\cite{a181224.01}
\begin{equation}
I_4=\frac{I_{\rm s}}{(1+r_1/r_2)(1- r_2 r_1)}\left[\dfrac{l_{\rm g}\eta_{\rm c} P_{\rm in}}{I_{\rm s} V}  - \ln\frac{1}{r_2 r_1}\right],
\end{equation}
where $r_1\equiv \sqrt{\mathcal{R}_1}$, $r_1\equiv \sqrt{\mathcal{R}_2}$, $l_{\rm g}$ is the thickness of the gain medium, $\eta_{\rm c}$ is the combined pumping efficiency, $I_{\rm s}$ is the saturation intensity, and $V$ is the volume of the gain medium. The saturation intensity is a constant related to the material of the gain medium and the frequency $\nu$ to be amplified, which is expressed as~\cite{a181218.01}
\begin{equation}
	I_{\rm s}=\dfrac{h\nu}{\sigma_{\rm s} \tau_{\rm f}},
\end{equation}
where $h$ is the Planck's constant, $\sigma_{\rm s}$ is the stimulated cross section area of the gain medium, and $\tau_{\rm f}$ is the fluorescence time of the gain medium. Since the fundamental beam has multiple transverse modes, it can be assumed having a homogeneous transverse intensity distribution. Therefore, we can compute the the fundamental beam power if we have obtained $I_4$; that is
\begin{equation}
P_\nu=I_4\pi a_{\rm g}^2.
\end{equation}

As depicted in Fig.~\ref{fig:model}, the equivalent reflectivity $\mathcal{R}_1$ is assumed equal to the loss factor of RR1 $\Gamma_{\rm RR1}$. Other loss factors are combined into $\mathcal{R}_2$; that is 
\begin{equation}
\mathcal{R}_2=\Gamma_{\rm g}^2\Gamma_{\rm air}^2\Gamma_{\rm RR2}\Gamma_{\rm diff},
\end{equation}
where $\Gamma_{\rm g}$, $\Gamma_{\rm air}$, $\Gamma_{\rm RR2}$, and $\Gamma_{\rm diff}$ are the loss factors of the gain medium, the air, the retroreflector RR2, and the diffraction of light field, respectively. The loss factor for laser to propagate in the air is modeled as~\cite{a200427.04}
\begin{equation}
\Gamma_{\rm air}(d)=\mathrm{e}^{-\alpha_{\rm air} d},
\end{equation}
where $\alpha_{\rm air}$ is the loss coefficient of the air which is related to the air condition. For clear air, $\alpha_{\rm air}=10^{-4}$~m$^{-1}$. The diffraction loss comes from beam propagation. Namely, due to diffraction, some of the light field is cast to the outside of the optical devices, resulting in energy loss. For paraxial condition  demonstrated in Fig.~\ref{fig:model}, the diffraction loss can be approximated by a closed-form formula proposed in~\cite{MXiong2021}.


\subsection{Second Harmonic Generation}
In this system, we employ an SHG crystal to generate the frequency-doubled beam. This process is based on the second-order nonlinear polarization of some birefringent crystals. In this process, two fundamental waves with the same frequency $\nu$ interact in the crystal; and then, the doubled frequency $2\nu$ is generated from the crystal.  Since the  fundamental beam passes through the SHG crystal twice in a round trip, the frequency-doubled beam can be expressed as
\begin{equation}
P_{2\nu}=2\eta_{\rm SHG}P_\nu,
\label{equ:shg}
\end{equation}
where $\eta_{\rm SHG}$ is the SHG efficiency. The SHG efficiency is strongly dependent on the incident angle of the fundamental beam. This requirement is called the phase matching. In our system, the beams inside RR1 are always perpendicular to M1, and therefore, perpendicular to the SHG crystal, ensuring a fixed incident angle. Hence, we can neglect the phase mismatch effects. Note that~\eqref{equ:shg} is true only when $\eta_{\rm SHG}$ is very small, as the effect to the fundamental resonance can be negligible under this condition. The SHG efficiency depends on the fundamental beam intensity at the SHG crystal which is expressed as $2P_\nu/[\pi w^2(0)]$. Here the fundamental beam intensity comprises the intensities of two oppositely traveling waves; and we assume they are equal at the SHG crystal under the low-SHG-efficiency condition.
The remaining factors that have effects on the SHG efficiency are the thickness of the crystal $l_s$ and the efficient nonlinear coefficient $d_{\rm eff}$.  Then, the SHG efficiency is approximated by~\cite{a181218.01}
\begin{equation}
\eta_{\rm SHG}=  \dfrac{8 \pi^2 d_{\rm eff}^2 l_{\rm s}^2 }{ \varepsilon_0 c \lambda^2 n_0^3}\cdot \dfrac{2P_\nu}{\pi w^2(0)},
\end{equation}
where $\epsilon_0$ is the vacuum permeability, $c$ is the light speed, $\lambda=c/\nu$ is the wave length of the fundamental beam, and $n_0$ is the refractive index of the SHG crystal. Note that the above calculation is based on the plane-wave approximation which is valid only when the SHG crystal's thickness $l_{\rm s}$ is smaller than the Rayleigh length ($z_{\rm R}=\pi w^2_{00}(0)/\lambda$) of the input beam.

\subsection{Communication Channel}

The communication carrier is the frequency-doubled beam generated by the SHG crystal. The frequency-doubled beam passes through the gain medium, and is then modulated by the EOM with intensity modulation. After that, the modulated beam transfers through the air and is received by the detector. Given the input signal $s_{\rm i}(t)$, the signal at the output of a direct detector (DD) is expressed as
\begin{equation}
s_{\rm o}(t)= s_{\rm i}(t)P_{2\nu}*h_{\rm EOM}(t)*h_{\rm air}(t)*h_{\rm det}(t) +n(t),
\end{equation}
where $h_{\rm EOM}(t)$, $h_{\rm air}(t)$, $h_{\rm det}(t)$ are the impulse response function of the EOM, the air, and the detector, respectively; and $n(t)$ is the additive Gaussian white noise. 

\begin{figure}[t]
	\centering
	\includegraphics[width=3.0in]{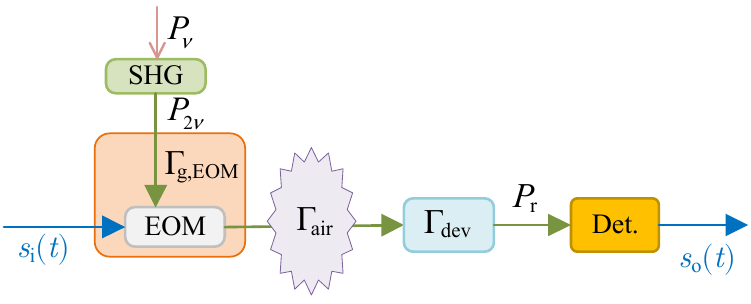}
	\caption{Communication channel model}
	\label{fig:channel}
\end{figure}

Since the carrier is light wave, it faces less distortion in the frequency domain when passing through these optical elements. Hence, we ignore the frequency-selective of the communication channel. The most important parameter with respect to the channel is its direct-current~(DC) gain. As depicted in Fig.~\ref{fig:channel}, during the transmission from the SHG crystal to the detector, the frequency-doubled beam is at first attenuated by the gain medium and the EOM due to the existence of impurities; then, some of the  beam energy is absorbed by the particles and vapors in the air; and finally, the detector receives the beam with some consumption due to reflection and diffraction. In addition, the transmitted beam also experiences attenuation when passing through the surface of these optical devices. Hence, the received optical power at the PD is expressed as
\begin{equation}
	P_{\rm r}=\Gamma_{\rm dev}\Gamma_{\rm det}\Gamma_{\rm air}\Gamma_{\rm g,EOM}P_{2\nu},
\end{equation}
where $\Gamma_{\rm g,EOM}$ is the loss factor of the combined body of the gain medium and the EOM, $\Gamma_{\rm det}$ is the loss factor of the PD, and $\Gamma_{\rm dev}$ is the combined loss factor that the frequency-doubled beam experiences when passing mirrors and lenses. 

The channel capacity can be computed according to Shannon's theory; that is~\cite{a200416.03, MXiong2021}
\begin{equation}
	\tilde{C}=\log_2\left\{1+\dfrac{(\gamma P_{\rm r})^2}{2e(\gamma P_{\rm r} + I_{\rm bk}) B+\frac{4kT B}{R_{\rm L}}}\right\},
\end{equation}
where $\gamma$ is the responsivity factor of the PD that converts light power into currents, $e$ is the electron charge, $I_{\rm bk} = 5100~\mu$A is the
 photocurrent induced by background light~\cite{a200427.02}, $B$ is the channel bandwidth, $k$ is the Boltzmann constant, $T$ is the temperature, and $R_{\rm L}$ is the load resistor.

\section{Numerical Results}
\label{sec:result}

	In this section, we present the parameter setting, and analyze the beam radius, the received power and the channel capacity of the proposed system, and finally make a comparison with the previous work. 
	
\begin{table} [t] 
	\caption{ Parameter Setting~\cite{a181218.01,a200527.01}}
	\renewcommand{\arraystretch}{1.2}
	\centering
	\begin{tabular}{ l l l}
		\hline
		\textbf{Parameter} & \textbf{Symbol} &  \textbf{Value} \\
		\hline
		Focal length of the lens & $f$ & $3$~cm\\
		Interval between the lens and mirror& $l$ & $3.015$~cm\\
		Stimulated emission cross section & $\sigma_{\rm s}$ & $15.6\times10^{-23}$ m$^{2}$\\
		Fluorescence lifetime&$\tau_{\rm f}$ & 100 $\upmu$s\\
		Fundamental beam wavelength & $\lambda$ & $1064$ nm\\
		Radius of gain medium aperture  & $a_{\rm g}$ & $3$ mm \\
		Gain medium thickness & $l_{\rm g}$ & $1$ mm\\
		Combined pumping efficiency& $\eta_{\rm c}$ & $43.9\%$\\
		Efficient nonlinear coefficient & $d_{\rm eff}$ & $4.7$~pm/V\\
		Refractive index&$n_0$ & 2.23 \\
		SHG medium thickness & $l_{\rm s}$ & $0.2$ mm\\
		PD's responsivity& $\gamma$ & $0.6$~A/W\\
		Temperature& $T$ & $298$~K \\
		Load resistor & $R_{\rm L}$& $10$~k$\Omega$\\
		\hline
	\end{tabular}
	\label{tab:paramReson}
\end{table}

\subsection{Parameter Setting}
	Many materials can be employed in the SSLR, including crystals and semiconductors. We employ Nd:YVO$_4$ crystal as the gain medium, as it has a large stimulated emission cross section $\sigma_{\rm s}$, which provides a high slope efficiency. This property makes it easy to initiate the resonance. Plus the high absorption coefficient to the pump power, the thickness of the crystal can be very small, i.e., in thin-disk shape. This shape of the gain medium meets the requirement of our system structure. However, rod-shaped materials can not provide a sufficient field of view (FOV). The parameters of this system are listed in Table~\ref{tab:paramReson}. Note that the combined pumping efficiency is comprised of a series of efficiency factors, including the quantum efficiency $\eta_{\rm Q}=95\%$, the Stocks factor $\eta_{\rm S}=76\%$, the overlap efficiency $\eta_{\rm B}=90\%$, the pump source efficiency $\eta_{\rm P}=75\%$, the pump source transfer efficiency $\eta_{\rm t}=99\%$, and the pump absorb efficiency $\eta_{\rm a}=91\%$~\cite{a181218.01,a200527.01}. Such parameter setting is under the assumption of multiple-transverse-mode operation. In our system, since the apertures of the optical devices are much bigger than the cross section of the fundamental mode, multiple modes are included in the resonance.

	Here we assume that the optical devices are with ideal fabrication so that no impurity is included. Therefore, we only consider that their loss factors come from reflection and transmission at the surface of them. Even though the surfaces of the lenses are coated with AR coating and the surfaces of mirrors are coated with HR coating, they can not exhibit $100~\%$ transmittance or reflectivity. In this numerical example, we set the power loss at each surface as $0.5\%$, which is close to the products that we can acquire in the market. Consequently, for the propagation of the fundamental beam, we set $\Gamma_{\rm RR1}=95.6\%$, $\Gamma_{\rm g}=98.5\%$, and $\Gamma_{\rm RR2}=97.5\%$; for the propagation of the frequency-doubled beam, we set  $\Gamma_{\rm dev}=96\%$,   $\Gamma_{\rm g,EOM}=97.5\%$,  $\Gamma_{\rm det}=99\%$.  Since the bandwidth is limited by the modulator, we set the bandwidth $B=800$~MHz, which is available as reported in~\cite{a200520.01}.
	
\begin{figure}[t]
	\centering
	\includegraphics[width=3.5in]{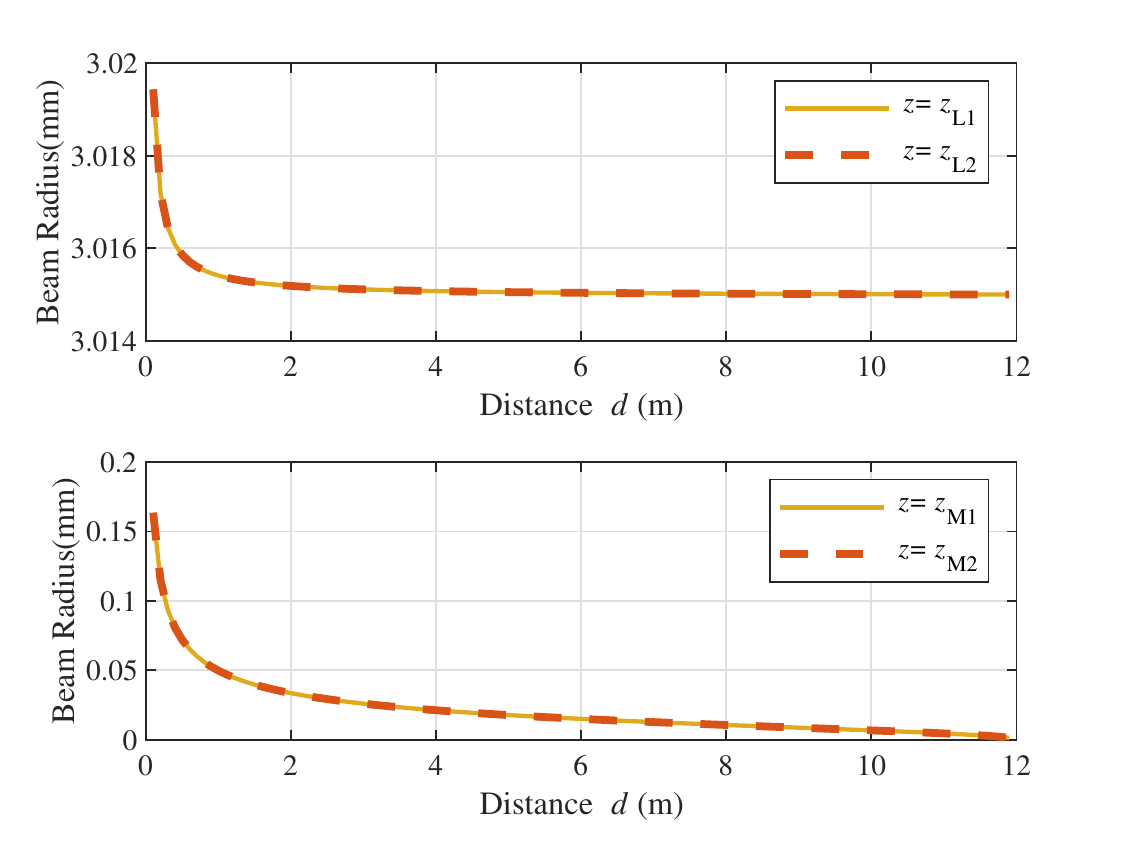}
	\caption{The relationship between the resonant beam radius and the  distance ($z_{\rm M1}$, $z_{\rm M2}$, $z_{\rm L1}$, and $z_{\rm L2}$ are the location on $z$-axis of the devices M1, M2, L1, and L2)}
	\label{fig:r-d}
\end{figure}

\subsection{Beam Radius}
	Fig.~\ref{fig:r-d} depicts the variation of the fundamental beam radius at the devices M1, M2, L1, and L2 with the increase of the transmission distance $d$. We can see that the beam radii at the mirrors M1 and M2 are consistent, and also the beam radii at the lenses L1 and L2 are consistent. The beam radii at the lenses go down rapidly when the distance starts to increase from $0$. Then, the beam radii become stable and approach $3.015$~mm 
as the distance continues to increase. Here we focus on the radius of the fundamental beam that incidents the SHG crystal. As mentioned before and shown in Fig.~\ref{fig:r-z}, the beam radius close to the mirror exhibits less changes. As a consequence, we can assume that the beam radius inside the SHG crystal is equal to that at the mirror. As illustrated in Fig.~\ref{fig:r-d}, the beam radius at the mirror decreases quickly as the distance increases, and then gradually decreases to $0$. The beam radius at the mirror is smaller than that at the lens. Therefore, the SHG efficiency is high although the crystal's thickness  is very small. 

\subsection{Performance Evaluation}

\begin{figure}[t]
	\centering
	\includegraphics[width=3.4in]{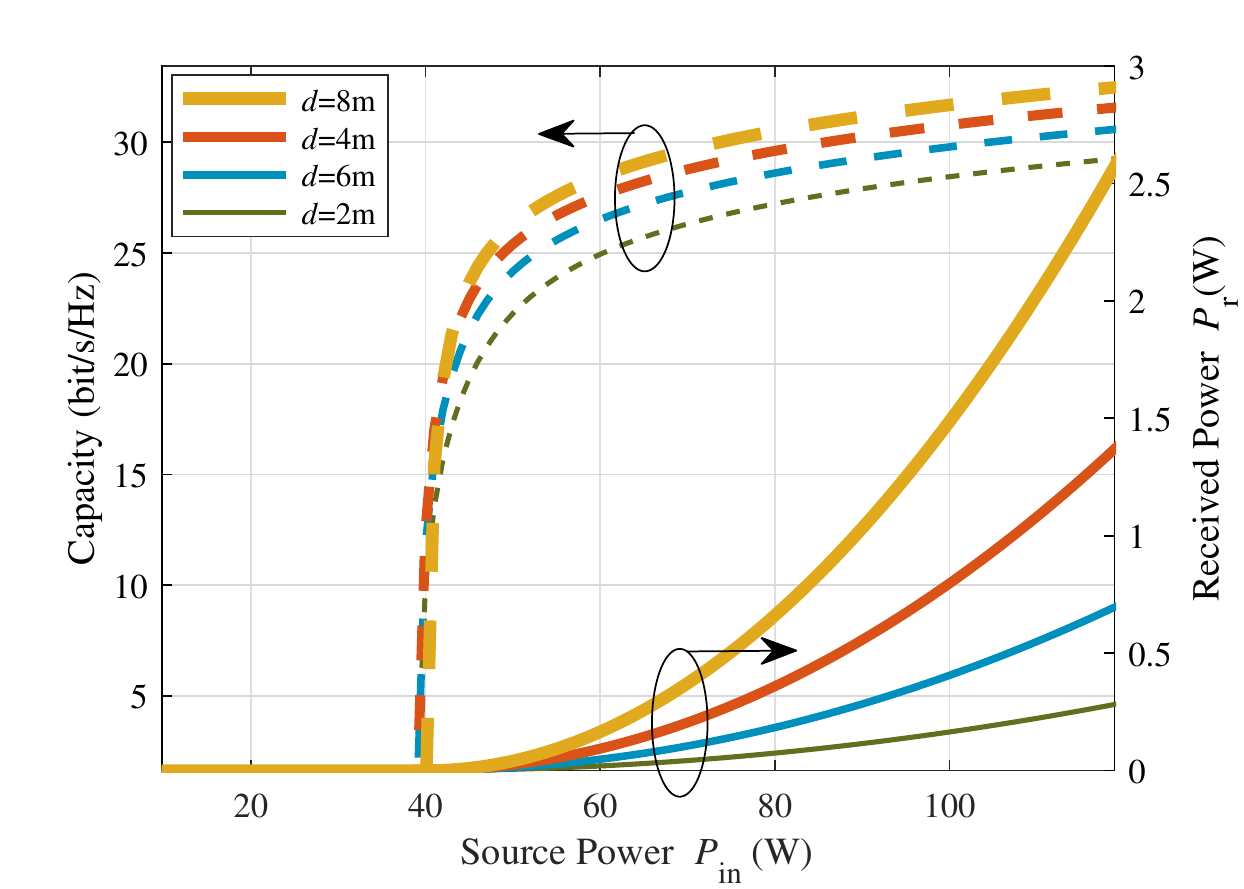}
	\caption{Capacity and received power \textit{vs.} source power (dashed lines: capacity; solid lines: received power;
		$d$: transmission distance)}
	\label{fig:Pow-Cap-Pin}
\end{figure}

\begin{figure}
		\centering
		\includegraphics[width=3.4in]{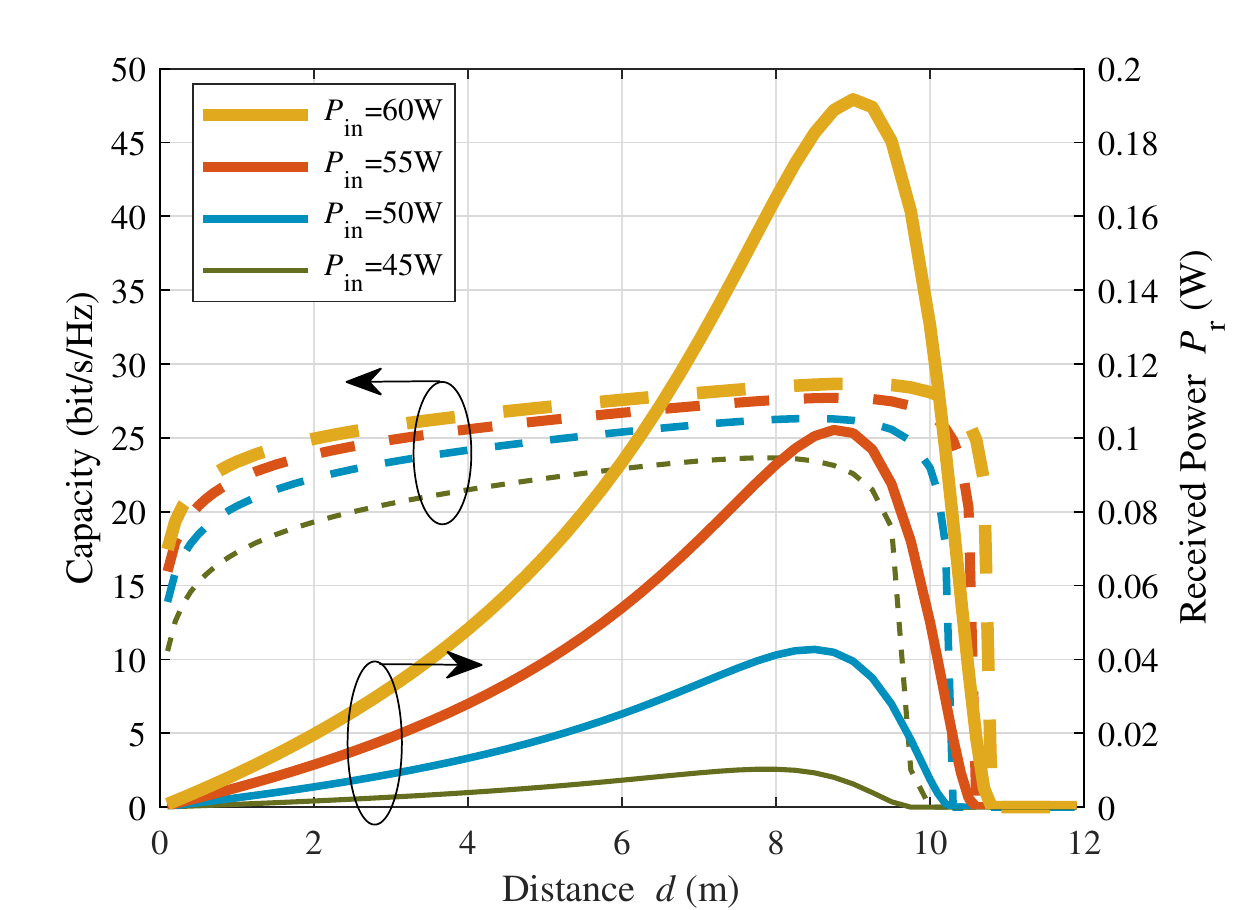}
		\caption{Capacity and received power \textit{vs.} transmission distance (dashed lines: capacity; solid lines: received power; $P_{\rm in}$: source power)}
		\label{fig:Pow-Cap-d}
	\end{figure}

\begin{figure}
		\centering
		\includegraphics[width=3.4in]{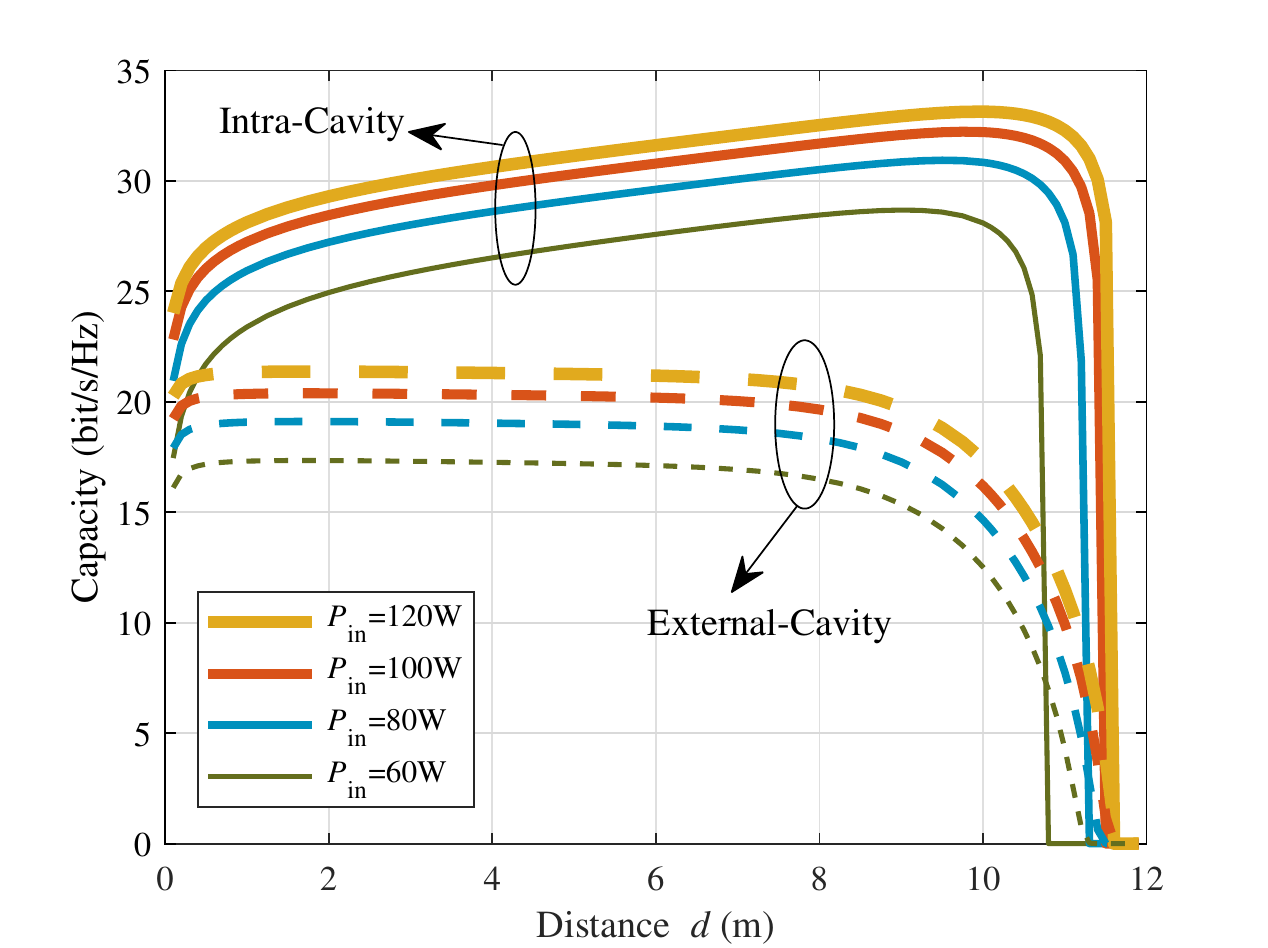}
		\caption{Capacity comparison between the intra-cavity SHG-based system and the external-cavity SHG-based system (solid lines: intra-cavity SHG; dashed lines: external-cavity SHG; $P_{\rm in}$: source power)} 
		\label{fig:CapCmp-d}
	\end{figure}

	We now present the numerical result with respect to the received power and the capacity of the proposed system. Fig.~\ref{fig:Pow-Cap-Pin} shows  the variation of the received power and the capacity with the increase of the source power. Fig.~\ref{fig:Pow-Cap-d} elaborates the effects of the distance on these performance aspects. Finally, Fig.~\ref{fig:CapCmp-d} depicts the performance comparison between this work and the previous work.

	As depicted in Fig.~\ref{fig:Pow-Cap-Pin}, there is a threshold for the source power $P_{\rm in}$. Below the threshold, there is no power to be generated. Therefore, for communication purpose, the source power should be set above the threshold so that the resonance can be initiated. As $P_{\rm in}$ increases, so does the growth rate of $P_{\rm r}$. This result reflects the fact that higher intensity of the fundamental beam induces higher SHG efficiency. The curve of the capacity is different from the curve of $P_{\rm r}$, i.e., the capacity grows rapidly at first, and then its growth rate decreases quickly  as $P_{\rm in}$ continues to increase. This result shows that the capacity is large enough when $P_{\rm in}$ is above $43$~W, for $d=8$~m. Generally, higher capacity is not useful, since the modulation order should be very high, which is unpractical due to the limited resolution of the analog-to-digital converter~(ADC) at present. As a result, according to the system model proposed in this paper, we can plot the relationship between the  capacity and the source power, and then choose a proper value for the source power.
	
	In Fig.~\ref{fig:Pow-Cap-d}, we can see the variation of the received power with the increase of the distance. In the previous paragraph for Fig.~\ref{fig:Pow-Cap-Pin}, we already see that longer distance has more power. $8$~m is much better than $2$~m. This is counter intuitive. The reason is as follows. As shown in Fig,~\ref{fig:r-d}, the fundamental beam radius at the left mirror decreases with the increase of $d$, which leads to the increase of intensity, and so does the SHG efficiency. However, as the distance increases continuously, the diffraction loss becomes dominant in the channel, eventually causing the resonance to cease. For this reason, the capacity also increases with distance at first and goes down to $0$ at last. But, the capacity variation is small. 
	
	Fig.~\ref{fig:CapCmp-d} illustrates the performance comparison between the system proposed in \cite{MXiong2021} and the system presented in this work. In previous work, the fundamental beam is extracted partially from the intra-cavity resonant beam by a splitter. However, in this work, the fundamental beam is the intra-cavity resonant beam itself. The extracted beam can be viewed as a loss to the resonant beam, and has very low power compared with the intra-cavity resonant beam. As the extracted beam passes through the SHG crystal, only a small portion of the power is used to generate the frequency-doubled beam. Therefore, the energy conversion efficiency of the previous work is very low. In this work, the intra-cavity resonant beam power is very high, since there is no power loss caused by extracting. Hence, the SHG efficiency is high enough to provide an enhanced $P_{\rm r}$ and capacity, as shown in Fig.~\ref{fig:CapCmp-d}. It can be observed from Fig.~\ref{fig:Pow-Cap-Pin} and Fig.~\ref{fig:CapCmp-d} that, to obtain at least $20$-bit/s/Hz capacity at $8$-m distance, we only need about $43$-W source power, while the previous work  needs $120$~W. The energy consumption is reduced by $64.2\%$ in this work.

\section{Conclusions}
\label{sec:con}
In this paper, we proposed a mobile optical communication system based on the spatially separated laser resonator~(SSLR). To avoid the echo interference inside the SSLR, we designed an intra-cavity SHG structure to generate a frequency-doubled beam for signal modulation. We elaborated the system design and established the analytical model for this system. Using this model, we demonstrated the beam radius distribution along the cavity's optical axis, and illustrated the performance in terms of received power and channel capacity. Numerical results showed that the proposed system can reduce the energy consumption by $64.2\%$ compared with the previous external-cavity design. Future studies will be focused on the stability in movement, and the improvement of pumping efficiency and transmission efficiency.



%

%



\ifCLASSOPTIONcaptionsoff
  \newpage
\fi




\bibliographystyle{IEEETran}
\small
%
\bibliography{mybib}
%
%

%

%
%







\end{document}